\documentstyle[prd,aps,floats,preprint,tighten,epsf]{revtex}
\draft

\begin{document}

\preprint{
    \parbox{1.5in}{%
       PSU/TH/158 \\
       hep-ph/9504220
    }
}

\title{
     Measuring transverse spin correlations by
     4-particle correlations in $e^{+}e^{-}\to  2 \ {\rm jets}$
}

\author{Xavier Artru}
\address{
  Institut de Physique Nucl\'eaire de Lyon, \\
  IN2P3-CNRS et Universit\'e Claude Bernard, \\
  43, boulevard du 11 Novembre 1918, \\
  69622 Villeurbanne Cedex, \\
  France \\
}
\author{John Collins}
\address{
  Penn State University, 104 Davey Laboratory, \\
  University Park, PA 16802, \\
  U.S.A.
}

\date{April 18, 1995}

\maketitle

\begin{abstract}
   The azimuthal distribution of pairs of particles in a jet is
   sensitive to the transverse polarization of the quark initiating
   the jet, but with a sensitivity that involves a nonperturbative
   analyzing power.  We show in detail how to measure the analyzing
   power from 4-hadron correlations in $e^{+}e^{-}\to  2 \ {\rm jets}$.  We
   explain the combinations of particle flavor that are likely to
   give the biggest effect.
\end{abstract}



\section{Introduction}
\label{sec:intro}

A useful probe of QCD would be to use appropriate observables in
a jet to probe the polarization of the parton initiating the jet.
For the case of longitudinal polarization, Nachtmann\cite{Nacht}
(and also Efremov\cite{Efr78})%
\footnote{See also work by Donoghue, Dalitz, Goldstein and
     Marshall\cite{Don,DGM}.}
proposed a three particle correlation to measure the helicity of
a quark. This observable is now called handedness\cite{EMT}.

A simpler measurement works for measuring transverse spin.  As
shown by Collins, Heppelmann and Ladinsky\cite{trfrag}, and by
Collins\cite{trfr-sgl}, the azimuthal dependence of hadrons
around the jet axis is a measure of transverse polarization for
quark jets.%
\footnote{
   Rotational invariance provides no trivial relation between
   longitudinal and transverse polarization for jet
   fragmentation, unlike the case of particle decay. A rotation
   of the spin also rotates the jet axis; this is quite analogous
   to properties of the spin states of a massless particle. Thus
   there is no contradiction between the need for a 3-particle
   measurement for longitudinal polarization and a 2-particle
   measurement for transverse polarization.
}
This effect we call the ``sheared jet effect''\footnote{
   This same effect is called the ``Collins effect'' by some
   authors\cite{AK,TM,Artru-Czyzewski}.
}.
It depends however
on non-perturbative effects in quark fragmentation, but at the
leading twist level.  So it is important to measure the analyzing
power of the observable for transverse spin.  Refs.\
\onlinecite{trfrag,trfr-sgl} show that the sheared jet effect is
compatible with the symmetries of QCD and is leading twist.  This
last property is in contrast to many other
transverse-spin-dependent observables, for example, the
deep-inelastic structure function $g_{2}$, which gives asymmetries
of order $1/Q$.

If a significant analyzing power can be obtained, then many
interesting measurements of processes induced by transversely
polarized hadrons can be made.  For example, Nowak\cite{Nowak}
lays out an experimental program for electron-hadron and
hadron-hadron collisions.
See also the proposal for a polarized hadron gas jet target at
LEP---the HELP proposal\cite{HELP}---and work by
Kotzinian\cite{AK}, and by Tangerman and Mulders\cite{TM}.
Among other things, one would have a
measurement\cite{trfrag,trfr-sgl} of the distribution ($\Delta _{T}q$ or
$h_{1}$) of transversely polarized quarks\cite{RS} in a transversely
polarized proton.  This is the one kind of twist-2 parton density
that is not measured at all by ordinary inclusive deep inelastic
scattering.

In this paper we show how the transverse-spin fragmentation
function can be measured in unpolarized $e^{+}e^{-}$ annihilation.
Perturbative QCD predicts that the quark and antiquark have quite
well correlated transverse spins.  Hence a measurement of the
azimuthal correlations of four particles, two from each of 2
jets, suffices to obtain the square of the desired fragmentation
function(s).

Already the SLD experiment has made a measurement\cite{SLD} of
the handedness of jets caused by the non-zero helicity of quarks
at the $Z^{0}$, albeit with a null result.
In addition, Ref.\ \onlinecite{DGM} cites an unpublished ``energetic
search with JADE data'' that produced no evidence for handedness.
This is discouraging
for the prospects for jet polarimetry, such as we are discussing.
However, one could readily imagine that main spin-dependent
correlations involve a few specific hadrons in a jet.  Then there
will be substantial dilution in the measured correlation due to
``combinatorial backgrounds'', just as there is in measuring
resonance peaks.  So it is possible that the transverse spin
correlation, which is the subject of this paper, would be less
subject to combinatorial backgrounds, since it involves fewer
particles.

We will give a precise specification of the measurements to be
made, and in addition list some ways in which the 4 hadrons
might be chosen to maximize the asymmetry.

In Sect.~\ref{sec:theory} we summarize the theoretical formalism
on which our proposed measurements is based, and we give the leading
order approximations for the cross sections and the spin
correlations.  Then in Sect.~\ref{sec:meas}, we specify in detail
how the experimental measurements are to be made.  Suggestions as
to how to enhance the signal by suitable choices of the flavors
of the detected hadrons are given in Sect.~\ref{sec:flavor}.
Some constraints can be obtained on the size of the effect from
measurements of jet fragmentation in unpolarized deep-inelastic
scattering, as we explain in Sect.~\ref{sec:UnpolDIS}.  Finally,
our conclusions are given in Sect.~\ref{sec:concl}.


\section{Theory}
\label{sec:theory}

We are interested in the production of two
(or more) jets in $e^{+}e^{-}$ annihilation followed by the
fragmentation of the jets into the particles we choose to
observe.  To include spin effects, the density matrix for the
parton must be included.  At lowest order in $\alpha _{s}$ we just have
the process $e^{+}e^{-} \to  q \bar q$.  The cross section for this
process, complete with the final-state density matrix was given
by Chen et al.\cite{CGJJ}:
\begin{eqnarray}
   \frac {d\hat\sigma _{q}}{d\Omega _{q}} \,
   \rho _{\alpha \alpha ';\beta \beta '}
  &=& \frac {N_{c}\alpha ^{2}_{em}}{8s}
    \Biggl[
      \biggl\{
         \frac {1}{2} Q_{q}^{2} (1+\cos^{2}\theta )
\nonumber \\
   &&\qquad\qquad\qquad
         + \frac {1}{2} \chi _{2}
           \left[ (1+\cos^{2}\theta )
             (v_{q}^{2}+a_{q}^{2}) (v_{e}^{2}+a_{e}^{2})
             + 8v_{e}a_{e}v_{q}a_{q} \cos\theta
           \right]
\nonumber \\
   &&\qquad\qquad\qquad
          - Q_{q}\chi _{1}
          \left[ v_{e}v_{q} (1+\cos^{2}\theta ) + 2a_{e}a_{q}\cos\theta
\right]
      \biggr\}
      \left( {\bf 1}^{q} \otimes {\bf 1}^{\bar q}
             - \sigma _{z}^{q} \otimes \sigma _{\bar z}^{\bar q}
      \right)
\nonumber \\
   &&\
      + \biggl\{
           \chi _{2} \left[v_{q}a_{q} (v_{e}^{2}+a_{e}^{2}) (1+\cos^{2}\theta )
               + 2v_{e}a_{e} (v_{q}^{2}+a_{q}^{2}) \cos\theta
              \right]
\label{sigma.density}\\
   &&\qquad
      - Q_{q}\chi _{1}
        \left[
           a_{q}v_{e}(1+\cos^{2}\theta ) +2v_{q}a_{e}\cos\theta
         \right]
     \biggr\}
        \left( {\bf 1}^{q} \otimes \sigma _{\bar z}^{\bar q}
              - \sigma _{z}^{q} \otimes {\bf 1}^{\bar q}
        \right)
\nonumber \\
   &&\
     + \left\{
        \frac {1}{2} Q_{q}^{2}
        + \frac {1}{2} \chi _{2} (v_{e}^{2}+a_{e}^{2}) (v_{q}^{2}-a_{q}^{2})
        - Q_{q}\chi _{1} v_{e}v_{q}
      \right\}
      \sin^{2}\theta
      \left( \sigma _{x}^{q} \otimes \sigma _{\bar x}^{\bar q}
             + \sigma _{y}^{q} \otimes \sigma _{\bar y}^{\bar q}
       \right)
\nonumber \\
   &&\
     - Q_{q} \chi _{1} v_{e} a_{q}
       \frac {\Gamma _{Z}M_{Z}}{s-M_{Z}^{2}}  \sin^{2}\theta
       \left( \sigma _{x}^{q} \otimes \sigma _{\bar y}^{\bar q}
              - \sigma ^{q}_{y} \otimes \sigma _{\bar x}^{\bar q}
       \right)
  \Biggr] \ .
\nonumber
\end{eqnarray}
This represents a parton-level cross section times a density
matrix for the outgoing quark and antiquark.  $\alpha $ and $\beta $ are the
indices on the density matrix for the quark, while $\alpha '$ and $\beta '$
are the indices for the antiquark.

In the above formula, right-handed axes are chosen in the
center-of-mass frame: The $z$ axis is defined to be the direction
of the quark, and the electron is defined to have momentum $k^{\mu } =
(E, -E \sin \theta , 0, E \cos \theta )$.  Ref.\ \onlinecite{CGJJ} defines
coordinates that are convenient for the antiquark:
$(\bar x, \bar y, \bar z) = (x, -y, -z)$, as used in
Eq.\ (\ref{sigma.density}).
But the final formulae
in the present paper will generally be written in a rotationally
covariant form in terms of vectors referred to one fixed frame.

The lepton couplings are
$v_{e} = 4 \sin^{2}\theta _{W} - 1$ and $a_{e}=-1$.
Also, $Q_{q}$ is the charge of the quark $q$ (in units of $e$), and
the couplings of the quarks to the $Z$ are
$v_{u} = 1 - \frac {8}{3} \sin^{2}\theta _{W} $,
$v_{d} = v_{s} = -1 + \frac {4}{3} \sin^{2}\theta _{W} $,
$a_{u}=1$, and $a_{d} = a_{s} = -1 $.  $N_{c}=3$
is the number of colors of a quark.  In addition
\begin{eqnarray}
   \chi _{1} &=& \frac {1}{16\sin^{2}\theta _{W}\cos^{2}\theta _{W}}
          \frac {s(s-M_{Z}^{2})}{(s-M_{Z}^{2})^{2}+\Gamma _{Z}^{2}M_{Z}^{2}} \
,
\\
   \chi _{2} &=& \frac {1}{256\sin^{4}\theta _{W}\cos^{4}\theta _{W}}
          \frac {s^{2}}{(s-M_{Z}^{2})^{2}+\Gamma _{Z}^{2}M_{Z}^{2}} \ .
\end{eqnarray}

The cross section is obtained by taking the trace of
Eq.~(\ref{sigma.density}) with a unit matrix
${\bf 1}^{q} \otimes {\bf 1}^{\bar q}$:
\begin{eqnarray}
  \frac {d\hat\sigma _{q}}{d\Omega _{q}} &=& \frac {N_{c}\alpha ^{2}_{em}}{4s}
    \Biggl\{
         (1+\cos^{2}\theta )
         \left[ Q_{q}^{2}
                + \chi _{2} (v_{q}^{2}+a_{q}^{2}) (v_{e}^{2}+a_{e}^{2})
                - 2 Q_{q}\chi _{1} v_{e}v_{q}
         \right]
\nonumber \\
   &&\qquad\qquad
         + \cos\theta
           \big[ 8 \chi _{2} v_{e}a_{e}v_{q}a_{q}
                   - 4 Q_{q}\chi _{1} a_{e}a_{q}
          \big]
      \Biggr\} \ .
\label{sigma}
\end{eqnarray}

The transverse-spin-dependent part is conveniently expressed in
Cartesian coordinates by taking a trace with
$\sigma _{i}^{q} \otimes \sigma _{i}^{\bar q}$.
We have to be careful with the rotated indices for $\bar y = - y$.
For the cross section times the density matrix, we get
\begin{eqnarray}
      \frac {d\Delta \hat\sigma _{qxx}}{d\Omega _{q}}
  = - \frac {d\Delta \hat\sigma _{qyy}}{d\Omega _{q}}
  &=& \frac {N_{c}\alpha ^{2}_{em}}{4s}
      \left[
        Q_{q}^{2}
        + \chi _{2} (v_{e}^{2}+a_{e}^{2}) (v_{q}^{2}-a_{q}^{2})
        - 2 Q_{q}\chi _{1} v_{e}v_{q}
      \right]
      \sin^{2}\theta  \ ,
\label{deltasigmaxx} \\
  \frac {d\Delta \hat\sigma _{qxy}}{d\Omega _{q}}
  = \frac {d\Delta \hat\sigma _{qyx}}{d\Omega _{q}}
  &=& \frac {N_{c}\alpha ^{2}_{em}}{4s}
      8 Q_{q} \chi _{1} v_{e} a_{q}
      \frac {\Gamma _{Z}M_{Z}}{s-M_{Z}^{2}}  \sin^{2}\theta  \ .
\label{deltasigmaxy}
\end{eqnarray}
The $xy$ and $yx$ terms are substantially smaller than the
diagonal terms, since they only arise from photon-$Z$
interference and then only from the imaginary part of the
$Z$ propagator.

By coupling the quark and antiquark to fragmentation functions,
we obtain cross sections for inclusive hadron production.  For
this, we use the definitions given in Ref.\ \onlinecite{trfrag}
for a fragmentation function and an asymmetry function for
particle pairs in a jet. The fragmentation function, $d_{12/c}$,
specifies the distribution of particle pairs in the jet generated
by an unpolarized parton of flavor $c$. The asymmetry function,
$A_{12/c}$, specifies the azimuthal asymmetry of the particle pair
when the quark initiating the jet has transverse polarization.
In effect, $A_{12/c}$ represents an analyzing power for quark
transverse polarization.  In this paper, we will use a
differently normalized asymmetry function, ${\cal A}_{12/c}$ instead
of $A_{12/c}$; it will be normalized so that it can be considered an
analyzing power, lying between $-1$ and $+1$.

We label the two detected particles in one of the jets $J$ by $1$ and
$2$ (or $H_{1}$ and $H_{2}$), and the two detected particles in the other
jet $J'$ by $1'$ and $2'$.  For the purposes of the present
discussion we will assume that $J$ is the quark jet and $J'$ is
the antiquark jet.  But ultimately we will wish to symmetrize the
cross section, between the two jets, since we will not be able to
distinguish which jet is which experimentally.

For each hadron, we will define a fractional momentum $z$, and
for the pairs $H_{1}H_{2}$ and $H'_{1}H'_{2}$, we will define invariant masses
$M$ and $M'$.  We will also define vectors ${\bf r}_{\perp }$ and
${\bf r'}_{\perp }$ that specify the relative transverse momentum within each
pairs of hadrons.  See Eq.\ (\ref{rdef}) below. Thus the hadron
variables are $z_{1}$, $z_{2}$, $z'_{1}$, $z'_{2}$, ${\bf r}_{\perp }$,
and ${\bf r'}_{\perp }$.  The fragmentation functions $d_{12/q}$ and $A_{12/q}$
are functions of $z_{1}$, $z_{2}$ and $|{\bf r}_{\perp }|$, and similarly for
the quantities in jet $J'$.  They also depend on the flavors of
the hadrons and partons involved.  Note that the invariant mass
$M$ of a hadron pair can be expressed in terms of $z_{1}$, $z_{2}$ and
$|{\bf r}_{\perp }|$.

With the above definitions, we have the following formula for the
inclusive production of 4 hadrons, in the kinematic region
where 2 of the hadrons are associated with each jet:
\begin{eqnarray}
   \frac {d\sigma }{dz_{1} dz_{2} d^{2}{\bf r}_{\perp } dz'_{1} dz'_{2}
d^{2}{\bf r'}_{\perp } d\Omega _{J} }
   &=&
   \sum _{q}  K  d_{12/q}(z_{1}, z_{2}, |{\bf r}_{\perp }|)
          d_{1'2'/\bar q}(z'_{1}, z'_{2}, |{\bf r'}_{\perp }|)
\nonumber\\
 && \
 \left[
   1 + \cos^2\theta  + 2 \hat a_{fb} \cos \theta
   + \sin^{2}\theta  \hat a_{NN} {\cal A}_{12/q} {\cal A}_{1'2'/\bar q}
      \frac {{\bf r}_{\perp } {\cal R} {\bf r'}_{\perp }}{|{\bf r}_{\perp }| \,
|{\bf r'}_{\perp }|}
 \right]
\nonumber\\
\rlap{$+$ Term with $(H_{1},H_{2})$ coming from $\bar q$ and
       $(H'_{1},H'_{2})$from $q$, and $\theta $ replaced by $\pi -\theta $\ .
}
\hspace{52mm}
\label{sigma.correl}
\end{eqnarray}
Here, ${\cal R}$ represents a rotation of $\pi $ about an axis
perpendicular to the scattering plane, so that
\begin{eqnarray}
   {\bf r}_{\perp } {\cal R} {\bf r'}_{\perp } &=&
   {r}_{y} {r'}_{y} - {r}_{x} {r'}_{x}
\nonumber\\
       &=& {\bf r}_{\perp } \cdot  {\bf r'}_{\perp }
     - 2 \frac {{\bf r}_{\perp } \cdot  {\bf k}_{e} {\bf r'}_{\perp } \cdot
{\bf k}_{e}}{E^{2}} .
\label{rRr}
\end{eqnarray}
The coordinates in the first line are those in the frame defined
above, and in the second line ${\bf k}_{e}$ is the 3-momentum of the
incoming electron, and $E$ is its energy (all in the
overall center-of-mass).  Equivalently, we can write
\begin{equation}
    {\bf r}_{\perp } {\cal R} {\bf r'}_{\perp } =
    - |{\bf r}_{\perp }| |{\bf r'}_{\perp }| \cos (\psi -\psi ') ,
\label{rRr.psi}
\end{equation}
where $\psi $ and $\psi '$ are the azimuthal angles of ${\bf r}$  and
${\bf r'}$  about their parent's jet axes, measured anticlockwise
from the direction of transverse part of the momentum of the
incoming electron, when the parent jet is viewed head on.

In Eq.~(\ref{sigma.correl}), we have defined an analyzing power
${\cal A}_{12/c}$ to be $|{\bf r}_{\perp }| / M$ times the corresponding
function $A_{12/c}$ in \onlinecite{trfrag}.  This we will call the
analyzing power of the jet for transverse spin, and can range
between $-1$ and $+1$.  But it has to have a kinematic zero when
${\bf r}_{\perp } \to  0$.
The number density of hadrons is thus of the
form:
\begin{eqnarray}
   dN(q\uparrow\to H_{1}\,H_{2}) &=&
   dz_{1}\, dz_{2}\,
   d^{2}{\bf r}_{\perp }\,
   d_{12/q}(z_{1},z_{2},|{\bf r}_{\perp }|)\,
\nonumber\\
&&
   \left[\, 1 + {\cal A}_{12/q}(z_{1},z_{2},|{\bf r}_{\perp }|)\,
\frac {{\bf p}_{q} \times {\bf r}_{\perp }}{|{\bf p}_{q}| \, |{\bf r}_{\perp
}|}
         \cdot {\bf s}_{\perp q}\,
   \right].
\end{eqnarray}

We have abbreviated the formula for the cross section, and have
defined coefficients
\begin{eqnarray}
    K &=&  \frac {N_{c}\alpha ^{2}_{em}}{4s}
         \left[ Q_{q}^{2}
                + \chi _{2} (v_{q}^{2}+a_{q}^{2}) (v_{e}^{2}+a_{e}^{2})
                - 2 Q_{q}\chi _{1} v_{e}v_{q}
         \right]
\label{Kdef}
\\
   \hat a_{fb} &=&
   \frac {4 \chi _{2} v_{e}a_{e}v_{q}a_{q} - 2 Q_{q}\chi _{1}
a_{e}a_{q}}{Q_{q}^{2} + \chi _{2} (v_{q}^{2}+a_{q}^{2}) (v_{e}^{2}+a_{e}^{2}) -
2 Q_{q}\chi _{1} v_{e}v_{q}} ,
\label{afb.def}
\\
   \hat a_{NN} &=&
   \frac {- Q_{q}^{2} - \chi _{2} (v_{e}^{2}+a_{e}^{2}) (v_{q}^{2}-a_{q}^{2}) +
2 Q_{q}\chi _{1} v_{e}v_{q}}{Q_{q}^{2} + \chi _{2} (v_{q}^{2}+a_{q}^{2})
(v_{e}^{2}+a_{e}^{2}) - 2 Q_{q}\chi _{1} v_{e}v_{q}} .
\label{aNN.def}
\end{eqnarray}
Obviously, $\hat a_{fb}$ and $\hat a_{NN}$ are parton-level
coefficients related to the forward-backward asymmetry and to the
transverse spin correlation $A_{NN}$ defined in the standard manner.

We will discuss the phenomenological consequences in the next
section.


\section{How to measure the spin correlation}
\label{sec:meas}

We now explain in detail how to make measurements that would be
sensitive to the spin-dependent fragmentation functions resulting
from the sheared jet effect. Generally we expect the highest
correlations of the state of hadrons in a jet with the state of
the parent parton for leading hadrons, that is, for hadrons with
the largest $z$ values.  We will therefore indicate how to
exploit this expectation to best advantage.  Additionally, there
are a number of choices that have to be made in defining which
hadrons are to be involved, and we will indicate suitable
choices.

\paragraph{Select 2-jet events}
Consider events for $e^{+}e^{-}\to 2\ {\rm jets}$, with some criterion
for distinguishing events with only 2 jets from events with 3 or
more jets.  (The use of 2-jet events probably cleans up the
signal, and is probably the simplest way of doing a preliminary
study.  However, a measurement of the polarized fragmentation
strictly according to the QCD definition is probably best done by
just looking at a four-particle inclusive cross section, with
suitable cuts on the momenta and angles of the particles.)

\paragraph{Choose and label hadrons}
Label the jets ``jet $J$'' and ``jet $J'\,$''.  For definiteness,
choose $J$ to be the jet with polar angle $0<\theta <\pi /2$.%
\footnote{The symmetry of the cross section formulae under
   exchange of the jets implies that exchanging $J$ and $J'$ will
   not affect the measurements, particularly if we ignore the
   forward-backward asymmetry. }
{}From each jet, pick two hadrons, which we will label $1$, $2$,
$1'$ and $2'$ (or $H_{1},H_{2}, H'_{1},H'_{2}$). One obvious possibility is
to choose $H_{1}$ to be the leading hadron (highest $z$), and $H_{2}$
to be the next-to-leading hadron. Then define $z_{1}$ and $z_{2}$ to be
the fractional energies of these two hadrons in jet
$J$.  Similarly define $z'_{1}$ and $z'_{2}$ to be the corresponding
quantities in jet $J'$. We have four such variables satisfying $z_{1}>z_{2}$
and $z'_{1}>z'_{2}$.
\footnote{\label{fn:choice}There are other possibilities for
   choosing the labels of the hadrons, but this seems the
   simplest. Note that the angular correlation is proportional to
   Eq.~(\ref{rRr}). Although it is invariant under exchange of
   the two jets, it changes sign when the two hadrons in a pair
   are exchanged.  So a definite choice of $H_{1}$ and $H_{2}$, etc, is
   necessary. Another obvious possibility (among others) would be
   to choose hadron 1 to be a $\pi ^{+}$ and 2 to be a $\pi ^{-}$. The
   measured spin correlation will depend on the method of
   choosing the hadrons, since the analyzing power depends on the
   momenta and flavors of the hadrons. We will explore this issue
   further in Sect.~\ref{sec:flavor}. }

\paragraph{Suitable cuts on the $z$'s}
Since the sheared jet asymmetry reverses sign when hadrons 1 and
2 are exchanged, it will vanish for $z_{1}=z_{2}$, if we do
not distinguish the flavors of the hadrons.  So it will
be a good idea to impose a cut like $z_{1}/z_{2}, z'_{1}/z'_{2}>2$, for
each jet.\footnote{%
   If a restriction on the flavors of the hadrons is imposed, as
   in footnote \ref{fn:choice}, then it may be more convenient not
   to impose $z_1 > z_2$.  Alternatively, the cut in
   $z_1/z_2$ should be more involved. }
Moreover, the effect is likely to be strongest for large $z$,
i.e., in the valence region.  So the data should be binned by the
values of the $z$'s (but with large bins, if needed to get good
statistics). The simplest approach for a first analysis would be
to take only events with $z_{1}+z_{2}>0.5$ and $z'_{1}+z'_{2}>0.5$. The
effect of these cuts will be to select asymmetric large momentum
hadrons, for which the desired asymmetry is presumably largest.
{\em The precise values for the cuts can, of course, be adjusted
to get the best signal, and the numbers we have suggested are not
sacrosanct.}

\paragraph{Cuts on hadron flavor}
As a way of enhancing the signal, one could restrict the analysis
to events in which the leading particles $H_{1}$, $H'_{1}$ have
opposite charges.  In addition, we would require that the
particles within each pair $H_{1}H_{2}$ and $H'_{1}H'_{2}$ have opposite
charges. This is an example of the criteria to be explored more
fully in Sect.~\ref{sec:flavor}.  It uses the information that
the leading particles are mostly pions, and have flavor
correlated with the flavor of the initiating quark.  Moreover,
with this constraint on the charges of the hadrons, the two
fragmentation asymmetry coefficients in Eq.~(\ref{sigma.correl}),
$A_{12/q}$ and $A_{1'2'/\bar q}$ are equal to each other, by charge
conjugation invariance. Thus the sign of the azimuthal
correlation is
definite, and to the extent that all the hadrons are pions, there
is no possibility of a cancellation of the effect from summing
over different flavors of hadron that have opposite values of
$A_{12/q}$.

The next steps will enable us to define an azimuthal direction
for each hadron pair, about its jet axis.

\paragraph{Jet axes}
Define an axis ${\bf J}$ and ${\bf J'}$
for each of the 2 jets (in the overall
center-of-mass).  This can be done by any standard algorithm.
The vectors ${\bf J}$ and ${\bf J'}$ are chosen to be unit
vectors.

\paragraph{Event plane}
Define an event plane by the two vectors ${\bf J} - {\bf J'}$ and
the beam direction. (There is a certain amount of arbitrariness
here. Since the jets will never be precisely back-to-back, it is
necessary to make some choice as to what constitutes the event
plane.  Our definition is simple, and seems to be good enough.)
It is likely to be useful to define an acollinearity from the 2
jet axes, as defined above.  One definition of a two-jet event
would be that the acollinearity is less than some suitable value.

\paragraph{Kinematics of the hadron pairs}
Define $M$ and $M'$ to be the invariant mass of the 2 chosen
hadrons in each of the jets:
\begin{eqnarray}
   M^{2} &=& (p_{1}+p_{2})^{2},
\nonumber\\
   M'^{2} &=& (p'_{1}+p'_{2})^{2}.
\end{eqnarray}
For each of the two pairs of hadrons, define a relative
transverse momentum by
\begin{eqnarray}
   {\bf r}_{\perp } =
   \frac {z_{2} {\bf p}_{1} - z_{1} {\bf p}_{2}}{z_{1} + z_{2}} ,
\nonumber\\
   {\bf r'}_{\perp } =
   \frac {z'_{2} {\bf p'}_{1} - z'_{1} {\bf p'}_{2}}{z'_{1} + z'_{2}} .
\label{rdef}
\end{eqnarray}
These definitions are arranged so that the longitudinal part of
the hadron momenta, parallel to the jets, cancels, to leading
order in $E$.
The vector ${\bf r}_{\perp }/M$ is equal to a $\pi /2$ rotation of the
spatial part of the vector $\Sigma$ in the Collins, Heppelmann
and Ladinsky paper \cite{trfrag}, as explained there.  The
invariant masses $M$ and $M'$ can be expressed in terms of the
other variables:
\begin{equation}
 M = \frac {z_{1}+z_{2}}{\sqrt {z_{1} z_{2}}}
   \sqrt {r_{T}^{2}  +
      \frac {z_{2} m_{1}^{2} + z_{1} m_{2}^{2}}{z_{1} + z_{2}}} ,
\label{eq:M}
\end{equation}
Thus the independent variables for the two
hadrons in jet $J$ are $z_{1}$, $z_{2}$ and ${\bf r}_{\perp }$, and similarly
for $J'$.

\paragraph{Azimuthal angles}
To visualize the asymmetry, we define azimuthal angles $\psi $
and $\psi '$ for each of the
vectors ${\bf r}_{\perp }$, ${\bf r'}_{\perp }$. These are
azimuthal angles around the jet axis. For each of the jets,
define $\psi =0$ to be in the plane of the scattering, and for the
sake of definiteness to be in the direction of the incoming $e^{-}$
(rather than the incoming $e^{+}$). The azimuths $\psi $ and $\psi '$ are to
be measured {\em anticlockwise} when the jet is coming {\em
toward} the observer. Alternatively, to avoid any lack of clarity
in the definitions of the angles, note that the combination on
which the cross section depends, $\cos(\psi -\psi ')$ is expressed in
terms of scalar products by Eqs.~(\ref{rRr}) and (\ref{rRr.psi}).

\paragraph{Experimental plots}
To check the azimuthal dependence in Eq.\ (\ref{sigma.correl}),
plot the $\psi -\psi '$ distribution for several bins of the overall
polar angle $\theta $ of the event and of the invariant masses $M$ and
$M'$ (or of the transverse momenta). Plots in $\psi +\psi '$ as well as a
scatter plot in $\psi ,\psi '$ should be also made, to look for other
possible kinds of azimuthal correlations. To avoid low
statistics, one should not have too many bins.  But there are
important dependencies on these variables.  So one should explore
this.

\paragraph{$M$ and $M'$ dependence}
At least one should split each of $M$ and $M'$ into
``small'' and ``large'' (say, above and below 1 GeV).
It is
quite possible that the analyzing power of the fragmentation
comes from interference effects, in which case it would be
instructive to perform the analysis at the mass of the $\rho $
meson, and just off the resonance peak.
It would also be appropriate to
tune the cuts to get the maximum effect.  One may do better to
make the bins in the transverse momentum variables $|{\bf r}_{\perp }|$
and $|{\bf r'}_{\perp }|$.

\paragraph{}
Since there is a characteristic $\theta $ dependence of the spin
correlation, but not of the higher twist QCD effects that also
produce azimuthal dependence, it would appear useful to split $\theta $
into ranges say 0 to $\pi /4$ and $\pi /4$ to $\pi /2$.

\paragraph{Predicted azimuthal dependence}
The cross section has the form%
\footnote{This equation is written in the approximation that we
   neglect the small off diagonal terms in the spin dependence,
   Eq.~(\ref{deltasigmaxy}).  If we wanted to be more exact,
   there would be a small angle to be added to $\psi -\psi '$. }
\begin{equation}
   \frac {d\sigma }{d\cos\theta  \, d\psi  \, d\psi '}
   \propto  1 + \cos^{2}\theta
     + 2 A_{fb} \cos \theta
     - \hat a_{NN} {\cal A}_{12} {\cal A}_{1'2'}
       \sin^{2}\theta  \cos(\psi -\psi ').
\label{asym}
\end{equation}
The coefficients ${\cal A}_{12}$ and ${\cal A}_{1'2'}$
are non-perturbative.
The other coefficients can all be deduced from Eqs.\
(\ref{sigma.correl}), (\ref{afb.def}) and (\ref{aNN.def}).  For a
purely electromagnetic process, i.e., well below the $Z$ peak, we
have $A_{fb}=0$, and $\hat a_{NN} = -1$.  At the $Z$ peak,
\begin{equation}
    \hat a_{NN}(Z) = \frac {g_{A}^{2}-g_{V}^{2}}{g_{A}^{2}+g_{V}^{2}} .
\end{equation}
With one flavor of quark, the value of the coefficient $\hat a_{NN}$ is
$0.74$ and $0.35$ for up-type and down-type quarks,
respectively.  The measured value will be a weighted average of
these.  (Note the reversal of sign from the electromagnetic case:
axial couplings dominate at the $Z$.)
Normally one will ignore the forward-backward asymmetry, which
produces the $\cos \theta $ dependence; it can only be measured if one
separates the flavor of the jets.

It is ${\cal A}_{12}$ and ${\cal A}_{1'2'}$ that we
wish to obtain from the analysis.
These quantities are the coefficients for the transverse
spin dependence of fragmentation. They depend on the
fragmentation variables, but to a first approximation one could
set them equal to the same fixed number, which is to be measured
by experiment.  (The number will depend on the experimental
cuts.) Note that the formula gives a maximum at $\theta =\pi /2$ and is
zero at $\theta =0$.  Moreover, it is invariant under exchange of the
labels of the two jets, so that the choice in the
labeling of the jets will not matter.

\paragraph{Estimator of the azimuthal correlation coefficient}

In fact, experiment will measure
\begin{equation}
C = {
      \sum _{f}  d_{12/f} d_{1'2'/\bar f}
      \ \times \
      \hat a_{NN} {\cal A}_{12/f} {\cal A}_{1'2'/\bar f}
   \over
      \sum _{f}  d_{12/f} d_{1'2'/\bar f}
   }
\end{equation}
for various types of hadron pairs and various bins in the $z$'s and
${\bf r}_{\perp }$'s.
In the above equation, $f = u,\,d,\,s,\,\bar u,\,\bar d$ or $\bar s$.
(We will assume that  $c\,\bar c$ and $b\,\bar b$ events can be
anti-tagged.)
$C$ can be determined by the maximum likelihood method or by the
simple recipe\cite{Besset}
\begin{equation}
C = -{
      { \sum _{{\rm event\in bin}}
        {\sin^{2}\theta \over 1+\cos^{2}\theta } \cos(\psi -\psi ') }
   \over
      { \sum _{{\rm event\in bin}}
        \left[{\sin^{2}\theta \over 1+\cos^{2}\theta } \cos(\psi -\psi ')
        \right]^{2}
       }
   } .
\label{estimator}
\end{equation}
This formula has the advantage that acceptance corrections in $\theta $
cancel between numerator and denominator.

\paragraph{General systematics}
There are some general characteristics to observe:
\begin{itemize}
    \item There will always be a kinematic zero in the azimuthal
    asymmetries when either $M$ or $M'$ goes to $2m_{\pi }$, since then
    it will be impossible to define an azimuthal angle.
    (Equivalently, we have a zero when either ${\bf r}_{\perp } = 0$ or
    ${\bf r'}_{\perp } = 0$.

    \item The sheared jet effect will be approximately proportional
    to $1/|{\bf r}_{\perp }|$ when $|{\bf r}_{\perp }|$ gets larger
    than some hadronic
    scale---e.g., 1 GeV. (And similarly for $|{\bf r'}_{\perp }|$.)  This is
    because the hadrons will then come from different jets, from
    well separated partons. As a function of $M$, the sheared jet
    effect is probably maximum in the resonance region and
    decreases at larger $M$.  It is not entirely obvious to us
    whether the governing scales are $|{\bf r}_{\perp }|$ and $|{\bf
    r'}_{\perp }|$ or $M$ and $M'$.

    \item There will be other higher twist effects that will
    correlate two jets.  These will be of order $M/Q$ or
    $|{\bf r}_{\perp }|/Q$. These effects should be small.
    For instance, the so-called ``string (or drag)
    effect''\cite{drag}
    is expected to grow with $MM'/Q^{2}$ for fixed
    $z_{i}$'s. Besides, it is independent of the orientation of the
    $e^{+}\,e^{-}\to q\,\bar q$ scattering plane, therefore can only be a
    function of ${\bf r}_{\perp }\cdot {\bf r'}_{\perp }$
    or $\psi + \psi '$. By contrast, the
    transverse spin azimuthal correlation we are discussing has
    an opposite $M$ dependence and is a function of $\psi - \psi '$.
    Therefore the two kinds of correlation can be easily
    disentangled.

\end{itemize}

\paragraph{Monte-Carlo calculations}
These can be used to estimate the results of the string effect.
A minimum experimental program would be to compare the 4-particle
correlation defined above with the prediction of standard
Monte-Carlo programs.  None of these have the sheared jet effect
built-in, but they typically do have the string effect.  When $M$
and $M'$ are reduced, say down to the $\rho $ mass, the $\psi $
dependence of the form given by (\ref{asym}) should be larger
than what is predicted by the Monte-Carlo.

\paragraph{Statistics needed}

To get an idea of the statistics required for a significant
signal, let us assume that the average sheared jet effect in one
particular bin and for one particular choice of hadrons is about
$A=0.3$ {\it per jet} and that the average $\hat a_{NN}$ is about
0.5. Then the infinite statistics value of the estimator $C$
defined by Eq.\ (\ref{estimator}) would be
\begin{equation}
    C_{{\rm true}} = \langle \hat a_{NN}\rangle  \times  A^{2} \sim 0.05.
\end{equation}

When the number of events, $N$, is large, and $C$ is small, the
variance of the estimator $C$
(Eq.\ (\ref{estimator})) satisfies
\begin{equation}
   \left\langle  \left(C_{{\rm measured}} - C_{{\rm true}}
          \right)^{2}
   \right\rangle
   = \frac {2}{N}
    \left\langle
        \left(\frac {\sin^{2}\theta }{1+\cos^{2}\theta }\right)^{2}
        \right\rangle ^{-1} .
\end{equation}
Taking
\begin{equation}
   \left\langle \left(
           {\sin^{2}\theta \over 1+\cos^{2}\theta }
        \right)^{2}
   \right\rangle  = {3 \over 4} \pi  - 2 \simeq 0.36,
\end{equation}
we see that about 2300 events are required in a particular bin,
in order to get a one $\sigma $ signal. This example shows that a rather large
number of events will have to be analysed, many tens of
thousands as a minimum.


\section{Role of flavor}
\label{sec:flavor}

In this section we explore how to choose the flavors of the
measured hadrons to optimize the chances of measuring a
significant effect.

\subsection{Theoretical prejudices}

Let us denote by
\begin{equation}
   \sigma\{ H'_1\,,H'_2\,;\,H_2\,,H_1\}
\end{equation}
the differential cross section of the semi-inclusive reaction
\begin{equation}
   e^+\,e^- \to H_1 + H_2 + \hbox{remainder of jet $J$} \quad + \quad
    H'_1 + H'_2 + \hbox{remainder of jet $J'$} .
\end{equation}
We will require the two measured hadrons in each jet to be the
highest energy hadrons in the jet, i.e., to be the leading
hadrons.

\begin{figure}
   \begin{center}
      \leavevmode
      \epsfbox{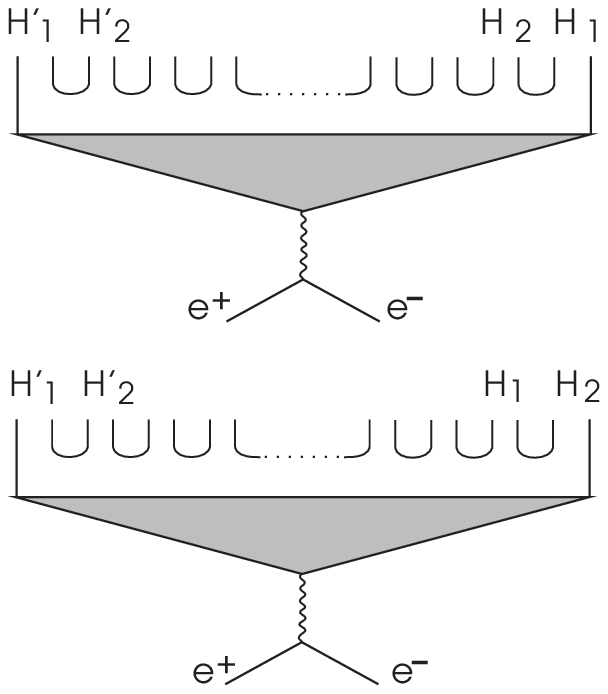}
   \end{center}
\caption{Schematic representation of fragmentation of jets.}
\label{fig:frag}
\end{figure}

Our analysis will proceed by considering the consequences of
a recursive fragmentation model of the string type or Feynman-Field
type.  The structure of such a model is symbolized by the
quark diagrams in Fig.~\ref{fig:frag}.  We then define the
``rank'' of a generated hadron in such a model to be 1 for the
endmost hadrons, 2 for the next-to-the-end hadrons, etc.  The
rank of a hadron in a jet is correlated with the
longitudinal momentum of the hadron, which may be measured by the
fractional momentum $z$ of the hadron or by its rapidity relative
to the jet axis.  Thus the leading hadron, the one with the
largest $z$, is likely to be the hadron of rank 1, etc.
The rank of a particle is a property of the method of generation
of the particle in a model.  Only the property of being leading
etc is observable.

Of course, these models are not exact.  But they do contain an
approximation to real QCD.  Moreover, hadron configurations that
do not result from such a model are suppressed.

A large fraction of the hadrons also comes from resonance decay.
Such a mechanism, which is not explicitly described by
Fig.~\ref{fig:frag}, would at first sight make the notion of
``rank" not very relevant. However, it is quite possible that
interference between different resonating diagrams favors
anisotropic decay of the resonances and restores the correlation
between rapidity and rank. This at least the lesson from the Dual
Resonance Model.

Motivated by these considerations, we make the following
assumption:
That the most favorable choice of the flavors of $H_1$, $H_2$,
$H'_1$, $H'_2$
for observing azimuthal correlations,
is when these particles could be of rank one and two
(but not necessarily in the order indicated by the labels 1 and 2),
in these ``quark cascade" models.\footnote{
   We recall that $H_1$ and $H_2$ are defined to the
   kinematically leading and next-to-leading particles in jet $J$
   (here also, not necessarily in the order indicated by the
   labels 1 and 2); similarly for the primed hadrons}
That is,
the azimuthal correlation between ${\bf r}_{\perp } \equiv
(z_2\, {\bf p}_{1} - z_1\,{\bf p}_{2}) / (z_1+z_2)\ $ and
${\bf r'}_{\perp }$ is a maximum for charges and strangenesses
compatible with $H_1$ and $H_2$ being of rank $\le 2$ in jet $J$
and $H'_1$ and $H'_2$ being of rank $\le 2$ in jet $J'$ (in a
recursive fragmentation model such as the string model or the
Feynman-Field model).

For simplicity, let us consider an idealistic sample which
contains no $c\bar c$ and $b\bar b$ events. (One may approach this case
by suitable rejection criteria.)  We determine the favorable
configurations in three steps:
\begin{itemize}

   \item We determine the optimum flavor states for
   $(H_1+H_2+H'_1+H'_2)$ with no regard to which particle of each
   pair is the leading one.

   \item We determine which combinations need not have strange
   quark production in the fragmentation.

   \item We determine which are the preferred configurations in
   relative values of $z_{1}$ and $z_{2}$.

\end{itemize}

Given our assumptions, it is easy to see from Fig.\
\ref{fig:frag} that
\begin{itemize}

   \item $(H_1+H_2)$ and $(H'_1+H'_2)$ must have nonexotic quantum numbers,
   {\it i.e.,} charges, strangenesses and third component of isospin
   between $+1$ and $-1$,

   \item $(H_1+H_2+H'_1+H'_2)$ must have nonexotic quantum
   numbers.

\end{itemize}
The first condition furthermore allows
$(H_1+H_2)$ and $(H'_1+H'_2)$ to be resonances. Then the fragmentation
amplitudes will readily acquire the phases necessary to get
nonzero asymmetries. If we apply these ``selection rules" to the
measurements in Sect.~\ref{sec:meas}, then that will enhance the
probability that $H_1$ , $H_2$, $H'_1$, $H'_2$ are all of rank
$\le 2$, and consequently are best correlated in flavor (and, we
hope, spin) with the parton initiating the jet.

\subsection{Case that all particles can be identified}

Let us first be very optimistic and assume that the detectors can
identify the $K^\pm$'s the  $\pi^\pm$'s, the $\pi^0$'s and the
$K_S$'s, with little contamination of other particles.  For the
neutral kaons, we use the notation ``$K_S$'', since a normal
detector cannot distinguish a $K^0$ from a $\bar K^0$, while
$K_L$'s mostly decay too late to be distinguished.

Then the possible experimental cases which met the conditions for
nonexotic {\it charges} are shown in table
\ref{table:CC1}.  Each box for a set of flavors has up to four
symbols: $\bullet$ and $\circ$.  If there are no marks in the
box, then the combination of hadrons does not satisfy all the
conditions of nonexotic {\it strangeness} or {\it isospin}.  If a
box has one or more marks, then it satisfies these conditions.
The precise meaning of the two symbols and of their placement
will be explained later.

Note that $\{ H'_1\,,H'_2\,;\,H_2\,,H_1\} = \{A,B;C,D\}$
is equivalent to $\{ H'_1\,,H'_2\,;\,H_2\,,H_1\} = \{D,C,B,A\}$,
together with the interchange of ${\bf p}_1$ and ${\bf p'}_1$, etc.
Due to CP invariance, it is also equivalent to
$\{ H'_1\,,H'_2\,;\,H_2\,,H_1\} = \{\bar A,\bar B; \bar C, \bar D\}$,
with a change of ${\bf p}_1$ into $-{\bf p}_1$, etc.
Combining these two symmetries,
$\{A,B;C,D\}$ is equivalent to $\{\bar D,\bar C; \bar B, \bar A\}$,
with ${\bf p}_1 \leftrightarrow -{\bf p'}_1$, etc.
This last operation corresponds to a symmetry
about the second diagonal of the table.
Other equivalences are
\begin{equation}
  \{6,j\} \leftrightarrow \{7,j\} \quad (j = 1,\cdots 5), \qquad
  \{i,6\} \leftrightarrow \{i,7\} \quad (i = 1,\cdots 5), \qquad
  \{6,6\} \leftrightarrow \{7,7\} ,
\end{equation}
where $\{i,j\}$ stands for the box at
$\{{\rm line}\ i,\ {\rm column}\ j\}$,
lines being counted {\it upwards}.
We are left with 55 non empty nonequivalent boxes.

A more realistic hypothesis is that one cannot distinguish
charged pions from charged kaons. Denoting by $H^\pm$ either a
$\pi^\pm$ {\it or} a $K^\pm$, we get the simpler table
\ref{table:CC2}.

\subsection{Correlations between flavor and longitudinal momentum }

Here we show some inequalities between cross sections for
different pairs of leading hadrons.  These inequalities are a
consequence of recursive fragmentation models.  To verify them
experimentally would be a check of some of our assumptions.

\subsubsection{Flavor compensation within one jet}

Suppose that $A+B$ is not exotic but $\bar A+B$ is. Then we
expect that
\begin{equation}
   \sum_{C,D}\ \sigma\{A,B: C,D\} > \sum_{C,D}\ \sigma\{\bar A,B; C,D\},
\end{equation}
since this is an inclusive cross section for production of
$\{A,B\}$.  The hadrons $A$ and $B$ are supposed to be the two
hadrons fairly close in rapidity, so that in a fragmentation
model depicted in Fig.~\ref{fig:frag}, they would be nearest
neighbors.
To get rid of some acceptance corrections, one may rewrite it as
\begin{equation}
{
  \sum_{C,D}\sigma\{A,B;C,D\} \times \sum_{C,D}\sigma\{\bar A,\bar B;C,D \}
\over
  \sum_{C,D}\sigma\{\bar A,B;C,D\} \times \sum_{C,D}\sigma\{A,\bar B;C,D \}
} > 1 .
\end{equation}
This inequality concerns lines or columns 4,5,6,7 of table \ref{table:CC1}
(line or column 4 of table \ref{table:CC2}).

\subsubsection{Flavor correlation between opposite jets}

Suppose that $D+C+A+B$ is not exotic but $D+C+\bar B+\bar A$ is. Then
\begin{equation}
   \sigma\{D,C;B,A\} > \sigma\{D,C;\bar B,\bar A\}\ ,\ {\rm or}
\end{equation}
\begin{equation}
{
  \sigma\{D,C;B,A\} \times \sigma\{\bar D,\bar C;\bar B,\bar A\}
\over
  \sigma\{D,C;\bar B,\bar A\} \times \sigma\{\bar D,\bar C;B,A\}
} > 1
\end{equation}
This inequality concerns the non-empty boxes of the $6\times 6$ sub-table
made by columns $6 \to 11$ and lines $6 \to 11$
in the upper right part of table \ref{table:CC1}.
(for table \ref{table:CC2}, it is the $2\times 2$
sub-table made by columns 5, 6 and lines 5, 6.)

In fact, these flavor
compensations within a jet and between jets are already known.
But it is worthwhile to check them again in the analyzed sample of events
(in an unpolarized fashion, {\it i.e.,} with an integration over azimuths).
If they are clearly seen, it will provide a confirmation that
our selection rules enhance the contributions
of rank $\le 2$ simultaneously in both pairs of mesons.
If they are barely seen, it indicates that contamination
from higher ranks may be too large to observe a significant
sheared jet asymmetry.
In that case, may be a more refined analysis,
for instance using the information about the three fastest particles
of each jet, may succeed.

\subsection{Dependence on momentum}

Each box is subdivided into 4 sub-boxes, according to the signs of
$z_1-z_2$ and $z'_1-z'_2$.
We make the following assignments:
Right upper sub-box: $H_1$ and $H'_1$ are both leading
(where ``leading'' means ``largest $z$ in the jet");
right lower: $H_1$ and $H'_2$ are leading;
left upper: $H_2$ and $H'_1$ are leading;
left lower: $H_2$ and $H'_2$ are leading.
These sub-boxes represent schematically the four quadrants of the
$(\,\log {z_1\over z_2}\,,\,\log {z'_1\over z'_2}\,)$ plane.

\begin{figure}
   \begin{center}
      \leavevmode
      \epsfxsize=0.4\hsize
      \epsfbox{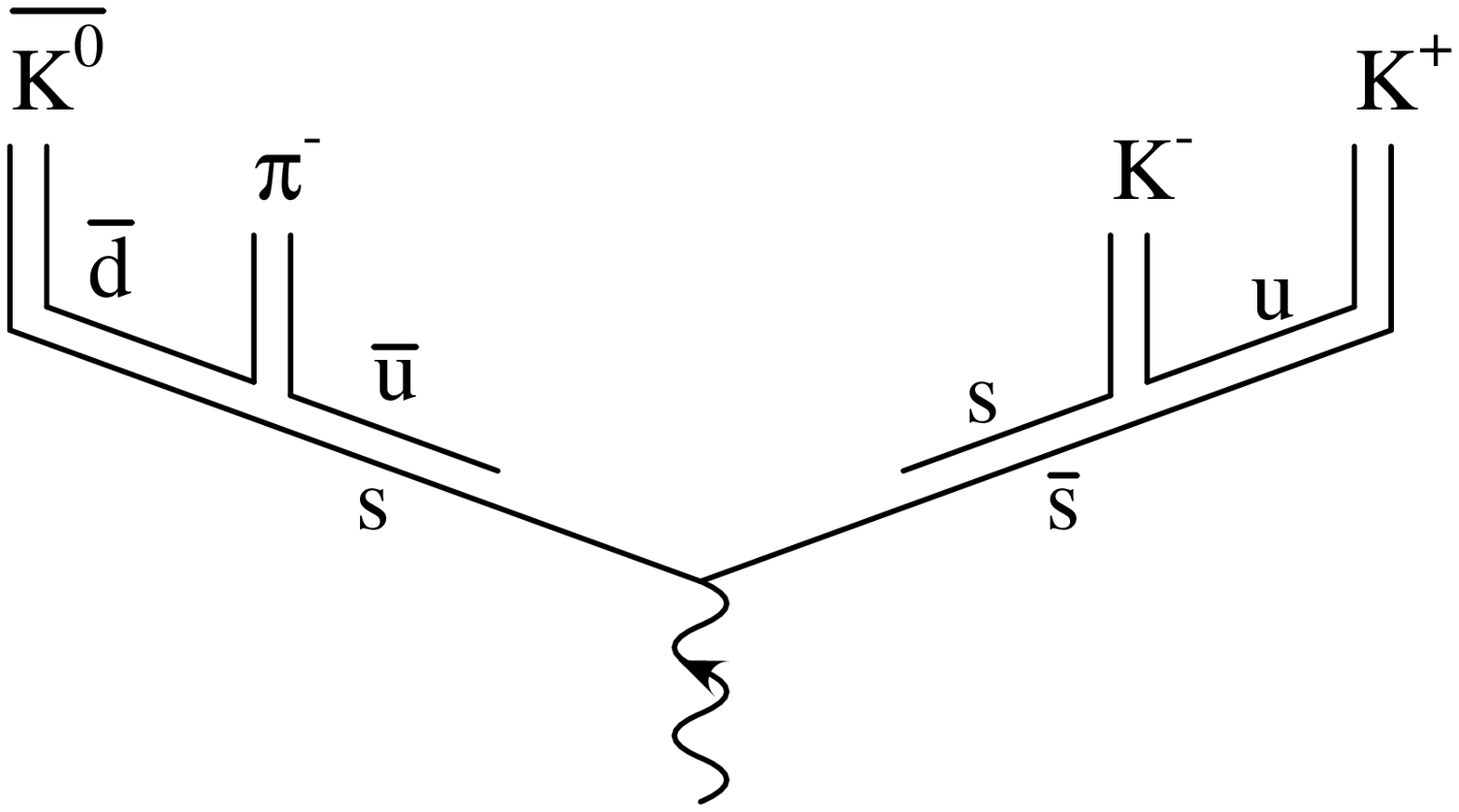}
      \hspace*{1cm}
      \epsfxsize=0.4\hsize
      \epsfbox{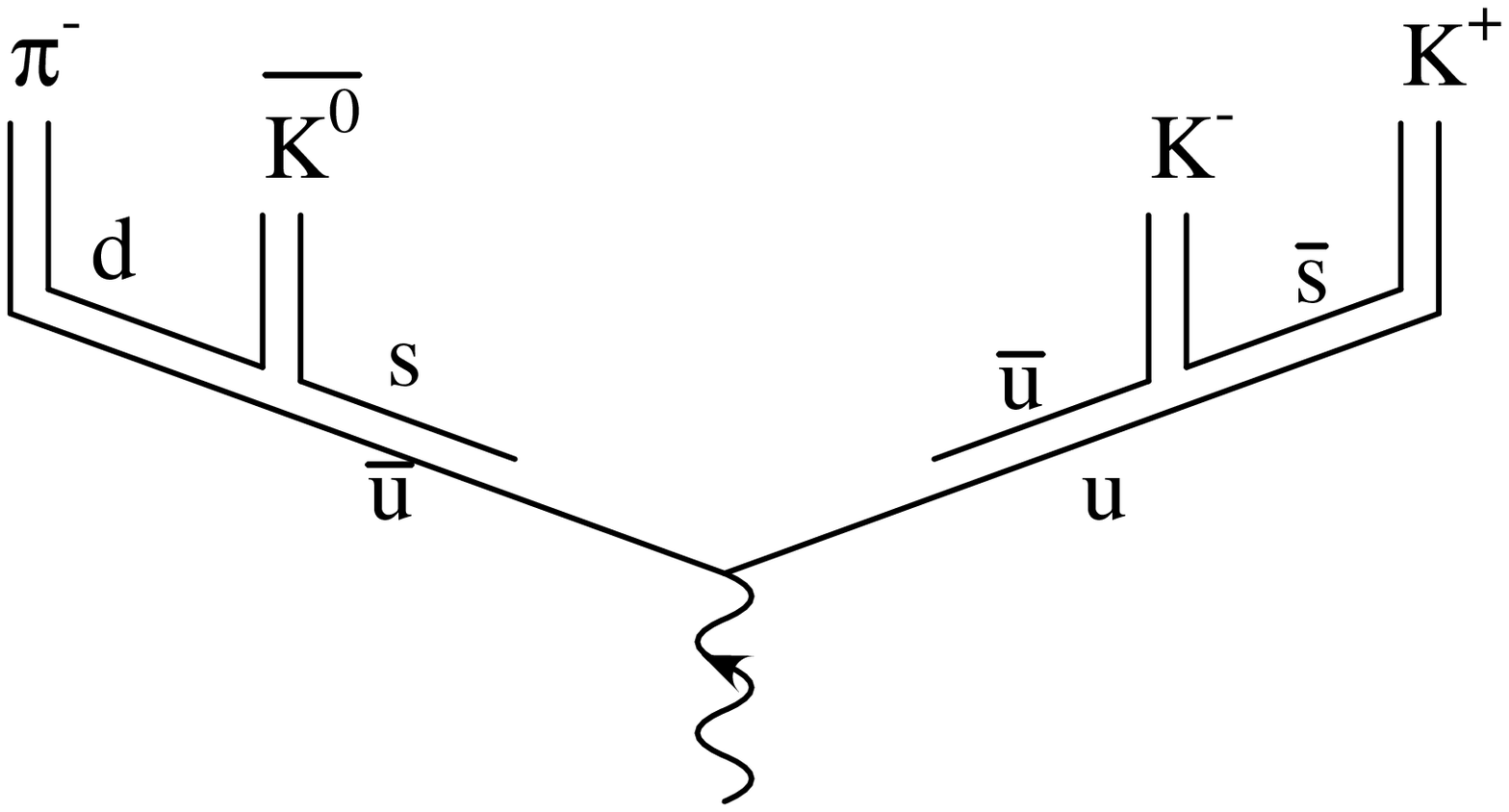}
   \end{center}
\caption{Quark diagrams for production of $K^{+}K^{-}\,\pi ^{-}\bar K^{0}$.
   The left-hand diagram corresponds to the lower right sub-box
   of box $\{4,9\}$) in table \protect\ref{table:CC1}, and the
   right-hand diagram corresponds to the upper right sub-box. The
   later ($\circ $) is less favored than the former ($\bullet$), due
   to the necessity of creating two $s\,\bar s$ pairs instead of
   one in the fragmentation process.
}
\label{fig:s.sbar.pair}
\end{figure}

It is clear from Fig.~\ref{fig:frag}
that the particle of first rank of jet $J$
and the particle of first rank of jet $J'$ make a non-exotic system.
Therefore sub-boxes where the
{\it two leading particles make a non-exotic system}
are favored ($\bullet$)
compared to sub-boxes (empty) where these particles make an exotic system.
The symbol $\circ$ means ``favored" but in a attenuated way, due to
the necessity of creating extra $s\,\bar s$ pairs relative to
other favored configurations in the same box.  An example is shown
in Fig.~\ref{fig:s.sbar.pair} for the case of the production of
$(H'_{1} H'_{2};\, H_{1} H_{2}) = (K^{+} K^{-};\, \pi ^{-} \bar K^{0})$.

As with the previous flavor correlations, the existence of
$z$-dependent favored configuration should be verified with
disregard of the polarization effects.

\subsection{Theoretical expectations about the size and the sign
            of the correlation}

In the ``string + $^3P_0$'' model \cite{Artru-Czyzewski}, the
single-meson sheared jet asymmetry essentially depends
on the rank of the meson: it is positive (in our convention) for odd ranks
and negative for even ranks.  Furthermore,
it decreases very fast when the rank
increases. Thus, for the 2-meson sheared jet asymmetry, the
fragmentation coefficient ${\cal A}$ is positive
if $H_1$ has rank 1, negative if $H_2$ has rank 1.
Of course, ranks cannot be determined experimentally,
but they are correlated with ordering in $z$, therefore this model predicts
that ${\cal A}$ has the same sign as $z_{1}-z_{2}$.
Accordingly, the azimuthal correlation function
${\cal A}_{12} {\cal A}_{1'2'}$
should have the following size and sign:
\begin{itemize}
\item large positive in a upper right or lower left sub-box endowed with
   a $\bullet$
\item large negative in a upper left or lower right sub-box endowed with
   a $\bullet$
\item for an empty sub-box,
    ${\cal A}_{12} {\cal A}_{1'2'}$ should at first sight
     be small and with no obvious sign.%
     \footnote{However, there will be a contamination from the sub-boxes
        which are not empty, because the lines
        $z_{1}=z_{2}$ and $z'_{1}=z'_{2}$ are not sharp boundaries
        between regions where the hadrons have particular ranks.}
     Thus, in the
     $(z_{1}-z_{2}, z'_{1}-z'_{2})$ quadrant corresponding to an empty
     sub-box, the sign of
     ${\cal A}_{12} {\cal A}_{1'2'}$ will be roughly determined
     by the nearest ``nonempty" quadrant. But the magnitude of
     ${\cal A}$ will decrease fast --- and its sign may change
     --- when one goes far from the $z_{1}=z_{2}$ and $z'_{1}=z'_{2}$
     lines.
\end{itemize}
Note that if the sheared jet asymmetry
depends on rank but with a sign opposite
to the one predicted by the ``string + $^3P_0$'' model, our conclusion
about the sign of ${\cal A}_{12} {\cal A}_{1'2'}$ remains unchanged.

Another prediction, which is absolute, concerns the case where we
choose $(1,2,1',2') = (+,-,-,+)$.  Then $(1,2)$ are
the antiparticles of $(1',2')$
${\cal A}_{12/q} = {\cal A}_{\bar 1\bar 2/\bar q}$. Hence we get a
definite sign for the correlation, since
${\cal A}_{12/q} {\cal A}_{\bar 1\bar 2/\bar q} =  {{\cal A}_{12/q}}^{2} \ge
0$.
Furthermore having $H_{1}= \bar H_{2}$ ensures that the correlation is
not suppressed by the zero of ${\cal A}_{12/q}$ at $z_{1}=z_{2}$ when 1
and 2 are identical particles.  This case is flagged by the large
bullets in table \ref{table:CC2} ({\large $\bullet$} instead of
$\bullet$).

\section{Unpolarized deep-inelastic scattering}
\label{sec:UnpolDIS}

Some constraints on the size of the sheared jet effect
can be obtained from
{unpolarized} deep-inelastic scattering, since measurements of the
angular distributions probe the angular momentum of the
hadron-pair.

Now, in a helicity basis relative to the jet axis, the
azimuthal distributions we are discussing are sensitive to
off-diagonal elements of this density matrix, and to those
elements with unit helicity flip: $\Delta h=\pm 1$. Hence upper bounds on
the off-diagonal elements can be obtained from the diagonal
elements.  Some of the diagonal elements can be measured in
unpolarized semi-inclusive deep-inelastic scattering.

Suppose one measures the angular distribution of
the pion pairs in their rest frame, as a function of the polar
angle $\theta _{\rho }$ from the jet axis.  This is a quantity that can in
fact be measured in jets of any kind (not only at LEP, but in
deep inelastic scattering, e.g., EMC and SMC).
The simplest case is when the pair comes from $\rho \to \pi \pi $.  Then
the general form of the distribution is
\begin{equation}
    f_{0} \sin^{2}\theta _{\rho } + f_{1} \cos^{2}\theta _{\rho },
\end{equation}
where $f_{0}$ and $f_{1}$ are the fractions of the mesons that are
produced in helicity 0 and helicity $\pm 1$ states.

First observe that a deviation from this form would be indicative
of interference effects.  Interference effects provide one
mechanism to obtain a sheared jet effect \cite{interference}.

More generally, recall that the helicity density matrix of a
transversely polarized quark is a linear combination of a unit
matrix and some off-diagonal terms which necessarily have
$\Delta h=\pm 1$.  Rotational invariance about the jet axis then implies
that the density matrix for the di-hadron in the final state of a
quark jet has the form of a diagonal matrix plus some $\Delta h=\pm 1$
terms.  The diagonal part is identical to when the quark is
unpolarized.  Diagonal terms with opposite helicity are equal

Consider the case of the $\rho $. Positivity of the eigenvalues of
the density matrix implies that a prerequisite for the
off-diagonal helicity flip $\Delta h=\pm 1$ terms, is that the density
matrix have terms of both helicity $0$ and $\pm 1$.

For example, suppose it happened that the angular distribution of
the two pions were close to
$\sin^{2}\theta _{\rho }$ or to $\cos^{2}\theta _{\rho }$.  Then the
$\rho $ would be close to pure helicity $0$ or to $\pm 1$
(respectively).  In that case the sheared jet effect would be small.
\footnote{Note that in the original version of
   \onlinecite{trfrag}, it was incorrectly stated that the
   sheared jet effect cannot exist at the $\rho $ peak in the absence of
   $\rho $-continuum interference.
}

Some data on this subject exist\cite{DISrho}, but not of a
sufficient precision.


\section{Conclusions}
\label{sec:concl}

The existence of spin-dependence in fragmentation would
provide a powerful tool for quark polarimetry, and would have a
substantial impact on the utilization experiments on polarized
high-energy scattering\cite{Nowak}.  Such measurements are
particularly important for transverse spin, where there is
otherwise a paucity of twist-2 observables.

Therefore we believe it is important to try and measure the
analyzing power of jets for quark transverse spin.  In this paper
we have explained in detail how to measure this quantity.  It
should be possible to do the measurements at current $e^{+}e^{-}$
colliders (SLC, LEP, TRISTAN).

The experimental quantity to be analyzed is a 4-hadron
correlation.  To enhance the effect, we have proposed a criteria
for the selection of the four hadrons in flavor and momentum.


\section*{Acknowledgements}

This work was supported in part by the U.S. Department of
Energy under grant DE-FG02-90ER-40577.
One of us (X.A.) wish to thank A. Kotzinian (HELP collaboration)
and W. Bonivento (DELPHI collaboration) for fruitful discussions.


\begin{table}


\def\B{$\bullet$}
\def\b{$\circ$}
\def\o{\phantom{$\bullet$}}
\def\pp{$\pi^+$}
\def\p-{$\pi^-$}
\def\po{$\pi^0$}
\def\kp{$K^+$}
\def\k-{$K^-$}
\def\ko{$K_S$}
\def\BB{\B\ \ \B}
\def\Bb{\B\ \ \b}
\def\Bo{\B\ \ \o}
\def\bB{\b\ \ \B}
\def\bb{\b\ \ \b}
\def\bo{\b\ \ \o}
\def\oB{\o\ \ \B}
\def\ob{\o\ \ \b}
\def\oo{\o\ \ \o}
\def\popo{$\pi^0\ \pi^0$}
\def\popp{$\pi^0\ \pi^+$}
\def\pmpp{$\pi^-\,\pi^+$}

\begin{center}
\begin{tabular}{c|c|c|c|c|c|c|c|c|c|c|c|c}
\multicolumn{1}{l}{}\\ 
\multicolumn{1}{c}{$H'_1$}  \\
\multicolumn{1}{c}{$H'_2$} &\multicolumn{5}{c}{}
                     &\multicolumn{1}{c}{$\rm H_2\,H_1$} \\
\multicolumn{1}{c}{$\downarrow$}  \\
\multicolumn{1}{l}{}\\ 
   & \ko\ko & \po\ko & \popo
   & \k-\kp & \pmpp & \k-\pp & \p-\kp
   & \ko\kp & \ko\pp & \po\kp & \popp & \\ \hline

\p- & \BB & \Bo & \BB & \oB & \oB & \oB & \oo & \BB & \oB & \Bo & \BB  &\p-  \\
\po & \BB & \Bo & \BB & \oB & \oB & \oB & \oo & \BB & \oB & \Bo & \BB  &\po  \\
\hline
\k- & \BB & \oB & \oo & \oB & \oo & \oo & \oB & \oo & \Bo & \oB & \oo  &\k-  \\
\po & \oo & \oo & \BB & \ob & \oB & \oo & \oo & \oB & \ob & \bo & \BB  &\po  \\
\hline
\p- & \oo & \oo & \BB & \ob & \oB & \oo & \oo & \oB & \ob & \bo & \BB  &\p-  \\
\ko & \BB & \oB & \oo & \oB & \oo & \oo & \oB & \oo & \Bo & \oB & \oo  &\ko  \\
\hline
\k- & \oo & \oo & \BB & \oB & \oB & \oo & \oo & \oB & \oB & \Bo & \BB  &\k-  \\
\ko & \BB & \Bo & \BB & \oo & \oB & \oB & \oo & \Bo & \oo & \oo & \BB  &\ko  \\
\hline
\k- & \BB & \oB & \oo & \oB & \oo & \oo & \oB & \oo & \Bo & \oB & \oo  &\k-  \\
\pp & \bb & \bo & \BB & \oo & \Bo & \oo & \bo & \oo & \oo & \oo & \oo  &\pp  \\
\hline
\p- & \bb & \bo & \BB & \oo & \oB & \ob & \oo & \Bo & \oo & \oo & \BB  &\p-  \\
\kp & \BB & \oB & \oo & \Bo & \oo & \Bo & \oo & \oo & \oo & \oo & \oo  &\kp  \\
\hline
\p- & \BB & \Bo & \BB & \oB & \oB & \oB & \oo & \BB & \oB & \Bo & \BB  &\p-  \\
\pp & \BB & \Bo & \BB & \Bo & \Bo & \oo & \Bo & \oo & \oo & \oo & \oo  &\pp  \\
\hline
\k- & \BB & \oB & \BB & \oB & \oB & \oo & \oB & \oB & \Bb & \bB & \BB  &\k-  \\
\kp & \BB & \oB & \BB & \Bo & \Bo & \Bo & \oo & \oo & \oo & \oo & \oo  &\kp  \\
\hline
\po & \BB & \Bo & \BB & \BB & \BB & \oB & \Bo & \BB & \oB & \Bo & \BB  &\po  \\
\po & \BB & \Bo & \BB & \BB & \BB & \oB & \Bo & \BB & \oB & \Bo & \BB  &\po  \\
\hline
\ko & \BB & \oB & \oo & \BB & \oo & \Bo & \oB & \oo & \Bo & \oB & \oo  &\ko  \\
\po & \bb & \bo & \BB & \oo & \BB & \ob & \bo & \Bo & \oo & \oo & \BB  &\po  \\
\hline
\ko & \BB & \bB & \BB & \BB & \BB & \Bb & \bB & \Bo & \Bo & \oB & \BB  &\ko  \\
\ko & \BB & \bB & \BB & \BB & \BB & \Bb & \bB & \Bo & \Bo & \oB & \BB  &\ko  \\
\hline
   & \ko\ko & \po\ko & \popo
   & \k-\kp & \pmpp & \k-\pp & \p-\kp
   & \ko\kp & \ko\pp & \po\kp & \popp & \\
\end{tabular}
\end{center}
\caption{Ratings of charge combinations.  In each square, a mark
(\B{} or \b) in one of the four positions means that there is a
quark diagram in which the corresponding hadrons are of rank 1,
and the other two hadrons are of rank 2.
For example, a mark in the upper right means that there is a
quark diagram where $H_{1}$ and $H_{2}$ are of ranks 1 and 2,
respectively, in one jet, while $H'_{1}$ and $H'_{2}$ are of ranks 1 and 2,
respectively, in the other jet.
A \b{} means that the hadrons require
production of an extra $s \bar s$ quark pair in the fragmentation
compared to other kinematic combinations in the same square.}
\label{table:CC1}
\end{table}

\begin{table}
\def\b{$\circ$}
\def\B{$\bullet$}
\def\BB{{\large $\bullet$}}
\def\ko{$K_S$}
\def\mp{$H^+$}
\def\m-{$H^-$}
\def\po{$\pi^0$}

\begin{center}
\begin{tabular}{c|rl|rl|rl|rl|rl|rl|c}
 & \ko & \ko & \po & \ko & \po & \po & \m- & \mp & \ko & \mp & \po & \mp  \\
\hline
\m- & \B & \B & \b & \B & \B & \B &    & \B & \B & \B & \B & \B & \m-  \\
\po & \B & \B & \b &    & \B & \B &    & \B & \B & \B & \B & \B & \po  \\
\hline
\m- &    &    &    &    & \B & \B &    & \B &    & \b & \B & \B & \m-  \\
\ko & \B & \B & \b & \B & \B & \B &    & \B & \B &    & \B & \B & \ko  \\
\hline
\m- & \B & \B & \B & \B & \B & \B &    & \BB& \B & \B & \B & \B & \m-  \\
\mp & \B & \B & \B & \B & \B & \B & \BB&    &    &    &    &    & \mp  \\
\hline
\po & \B & \B & \B &    & \B & \B & \B & \B & \B & \B & \B & \B & \po  \\
\po & \B & \B & \B &    & \B & \B & \B & \B & \B & \B & \B & \B & \po  \\
\hline
\ko & \B & \B &    & \B &    &    & \B & \B & \B &    &    & \B & \ko  \\
\po & \b & \b & \b &    & \B & \B & \B & \B & \b &    & \b & \b & \po  \\
\hline
\ko & \B & \B & \b & \B & \B & \B & \B & \B & \B &    & \B & \B & \ko  \\
\ko & \B & \B & \b & \B & \B & \B & \B & \B & \B &    & \B & \B & \ko  \\
\hline
 & \ko & \ko & \po & \ko & \po & \po & \m- & \mp &  \ko & \mp & \po & \mp  \\
\end{tabular}
\end{center}
\caption{Ratings of charge combinations, ignoring $\pi ^{\pm }-K^{\pm }$
   difference.}
\label{table:CC2}
\end{table}

\end{document}